\documentclass{iau} 
\usepackage{graphicx}
\usepackage{amsmath}
\usepackage{amsfonts}
\usepackage{amssymb}
\usepackage{siunitx}

\title[Adaptive dynamics in external galaxies] 
{Accreted Globular Clusters \\in External Galaxies: \\Why Adaptive Dynamics is not the Solution}

\author[Sophia Lilleengen \& Wilma Trick \& Glenn van de Ven]   
{Sophia Lilleengen$^{1,2,}$
 \and Wilma Trick$^3$ \and Glenn van de Ven$^{1,4}$}

\affiliation{$^1$European Southern Observatory, Karl-Schwarzschild-Str. 2, \\85748 Garching b. M\"unchen, Germany\\ email: {\tt sophialilleengen@gmail.com} \\[\affilskip]
$^2$Department of Physics and Astronomy, University of Heidelberg, \\Im Neuenheimer Feld 226, 69120 Heidelberg, Germany\\[\affilskip]
$^3$Max-Planck-Insitut f\"ur Astrophysik, Karl-Schwarzschild-Str. 1, \\85748 Garching b. M\"unchen, Germany\\[\affilskip]
$^4$ Department of Astrophysics, University of Vienna,\\T\"urkenschanzenstrasse 17, 1180 Wien, Austria}

\pubyear{2019}
\volume{353}  
\jname{Galactic Dynamics in the Era of Large Surveys}
\editors{M. Valluri \& J. Sellwood}
\begin{document}

\maketitle

\begin{abstract}
Many astrophysical and galaxy-scale cosmological problems require a well determined gravitational potential which is often modeled by observers under strong assumptions. Globular clusters (GCs) surrounding galaxies can be used as dynamical tracers of the luminous and dark matter distribution at large (kpc) scales. A natural assumption for modeling the gravitational potential is that GCs accreted in the same dwarf galaxy merger event move at the present time on similar orbits in the host galaxy and should therefore have similar actions. We investigate this idea in one realistic Milky Way like galaxy of the cosmological \textit{N}-body simulation suite \textit{Auriga}. We show how the actions of accreted stellar particles in the simulation evolve and that minimizing the standard deviation of GCs in action space, however, cannot constrain the true potential. This approach known as ‘adaptive dynamics’ does therefore not work for accreted GCs. 
\end{abstract}

\firstsection 
\section{Introduction}
One of the most fundamental galaxy properties is its total mass and how it is distributed. A galaxy's mass distribution give rise to a gravitational potential that governs how the objects move. While the mass of the Milky Way (MW) can be estimated from resolved dynamical tracers (e.g. \cite{BR13}), mass measurements of external galaxies rely mainly on the net velocities of unresolved stellar populations. 
In the MW, some of these dynamical tracers are stellar streams. Recent work by \cite{Sanderson15} (2015, 2017) suggest that actions of streams could, under the assumption of simple models, be useful to constrain the Galactic gravitational potential. Since in external galaxies stars cannot be resolved anymore, methods used on MW stellar streams could be adapted to accreted globular clusters (GCs) under the same assumptions. 

In this work, we test the idea of adaptive dynamics (\cite{Binney05}) in external galaxies. Adaptive dynamics suggests to use dynamical tracers on similar orbits, be it stellar streams in the halo or resonance moving groups in the disk, and to attempt to make these features as sharp as possible in action space, i.e. finding the potential in which the tracers are all approximately on the same orbit. We carry out the tests in a cosmological \textit{N}-body simulation of an \textit{Auriga} (\cite[Grand \etal\ 2017]{AurigaGrand17}) galaxy, where we have full 6D phase-space information for all particles.

\section{Adaptive dynamics in the \textit{Auriga} simulations}
{\underline{\it Simulation data}}. \textit{Auriga} is a suite of 40 cosmological magneto-hydrodynamical zoom simulations of the formation of galaxies in isolated MW mass dark halos including a galaxy formation physics model, AGN feedback and magnetic fields built with the moving mesh code \texttt{AREPO} (\cite{AREPO}). We carry out the tests in halo 24.

{\underline{\it Actions}}.
The orbit of an object can be described by either the time evolution of six dimensional position-velocity information or by a particular set of integrals of motion, the actions. Together with angle coordinates, they create a canonical coordinate system. Actions are calculated from an individual (x, v) measurement and the gravitational potential the tracers are moving in. For more details, we refer to \cite{BT08} and \cite{SandersBinney16}. We calculate actions using the galactic dynamics package \texttt{galpy} (\cite{Bovy15}).

{\underline{\it Potential model}}.
We describe the potential of the simulated disk galaxy with an analytic, axisymmetric, multi-component potential so that we treat these simulations as observers usually model external galaxies. The model consists of a Miyamoto Nagai disk, a Hernquist bulge for the stellar central spheroid and a NFW halo with in total 8 free parameters. As ground truth, we use a best-fit to the particle density distribution.

{\underline{\it Merger events}}.
We identify the three biggest merger events. All are minor mergers. The older mergers (prog4, 9.5, and prog3, 8.7 Gyr ago) have a merger ratio of 1:30 and the most recent merger (prog2, 3.2 Gyr ago) has a ratio of 1:50. Each accreted particle is treated as a GC due to the mass resolution of the stellar particles in \textit{Auriga} ($5\cdot10^4 ~M_\odot$).

{\underline{\it Adaptive dynamics}}. The GCs accreted by one merger are assumed to move on similar orbits. Particles on similar orbits have by definition similar actions. In the wrong potential, the GCs will not move on similar orbits and the spread in actions will be larger. By minimizing their spread with respect to the free potential parameters, we should be able to constrain the potential of the galaxy.

\begin{figure}[htbp]
\begin{center}
 \includegraphics[width=3.4in]{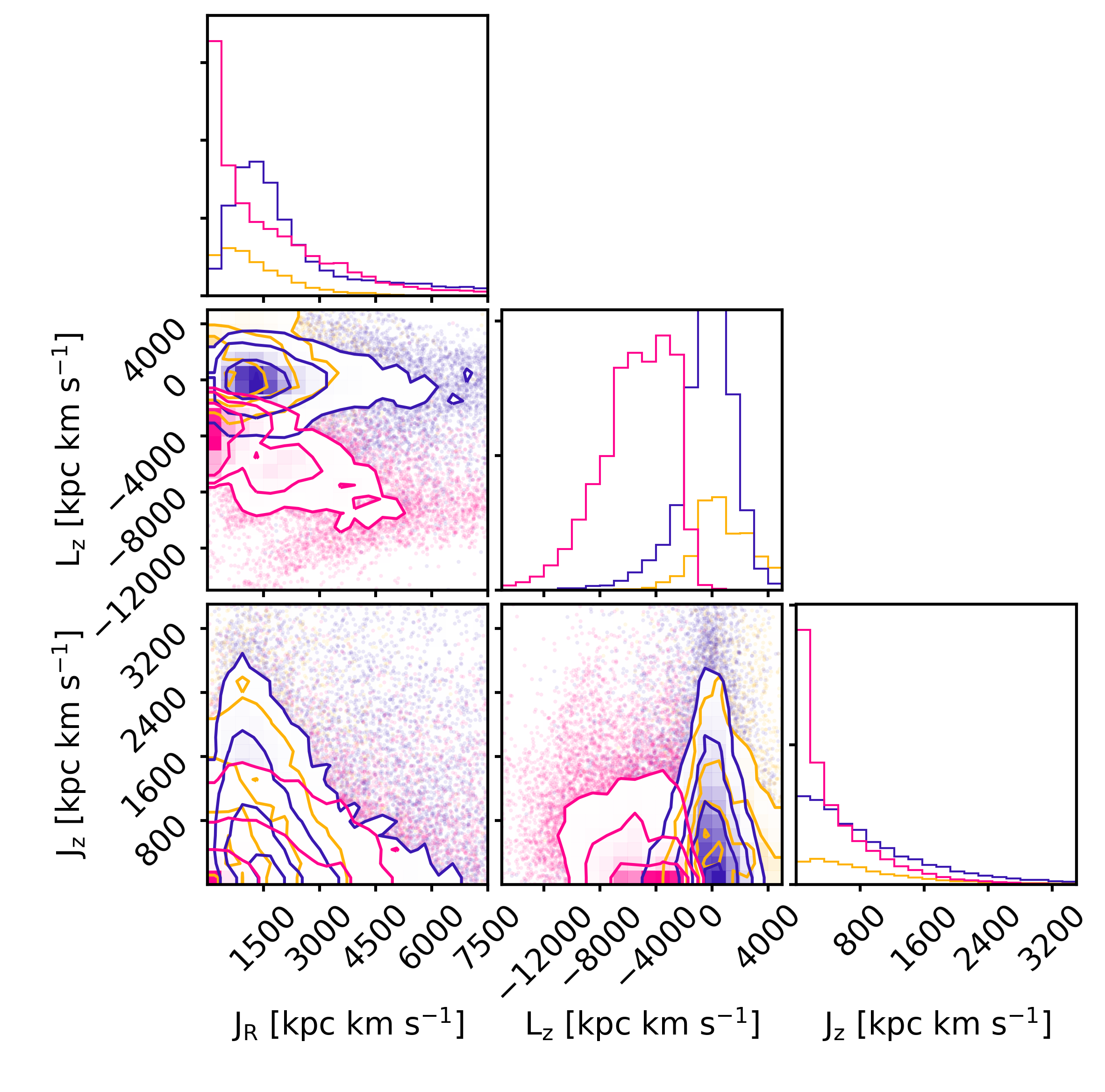} 
 \caption{Selected GCs from three different dwarf galaxy merger events in action space estimated in the "true" potential model. The diagonal subplots show histograms of each action and the other panels show 2D histograms of each action pair. The most recently accreted GCs (prog2, pink) can be clearly distinguished in $L_z$ and $J_R$ while the other merger remnants (prog3, blue and prog4, yellow) distribute around similar means. prog3 and prog4 have a mean angular momentum of approximately 0 kpc~km~s$^{-1}$  and mean higher radial action than prog2 indicating they are accreted in the halo while prog2 ended up in the disk and is counter-rotating. All groups have broad distributions in the actions.}
 \label{fig1}
\end{center}
\end{figure}

\begin{figure}[htbp]
\begin{center}
 \includegraphics[width=4.8in]{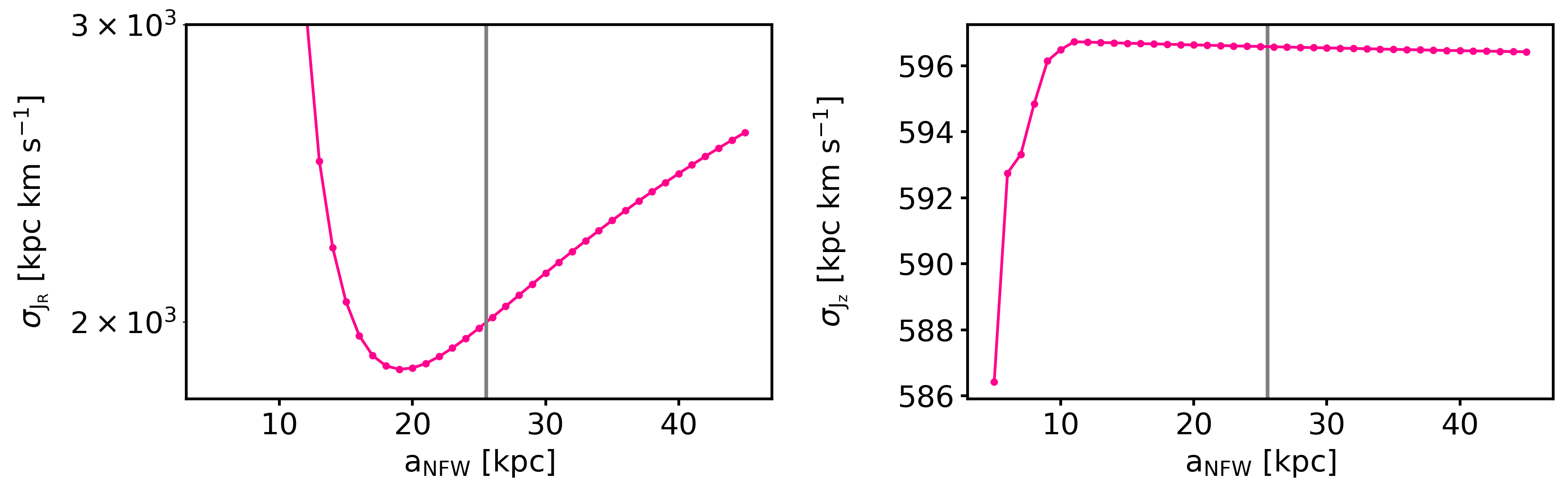} 
 \caption{Standard deviations of the radial action $J_R$ and vertical action $J_z$ in potentials with varying scale length $a_{NFW}$ of the dark halo while all other parameters are fixed at their "true" value. We would expect the standard deviation to be mimized at the "true" potential which is indicated with the vertical grey line at $a_{NFW} = 25.2$~kpc. The minimum of the radial action is, however, at $a_{NFW} = 19$~kpc. The vertical actions would prefer very small or higher scale lengths than the "true" value, but the differentiating power in $\sigma_{J_z}$ is much smaller than in $\sigma_{J_R}$.}
 \label{fig2}
\end{center}
\end{figure}

{\underline{\it Testing adaptive dynamics}}. In Figure \ref{fig1}, we plot the accreted GCs of the merger events in action space. Prog4 is the most compact group, while especially prog3 is very dispersed. None of the distributions looks like a $\delta$-function. We quantify the compactness of the GC merger groups by measuring the standard deviation of the corresponding action distribution. To test adaptive dynamics, we vary only one potential parameter, the scale length of the dark matter halo and calculate for each potential and progenitor the action distribution and its standard deviation.
Figure \ref{fig2} shows the change of the standard deviations in the radial and vertical action with the variation of the free potential parameter. They do not have their minimum at the "true" potential but we see that vertical and radial actions prefer different potentials. This leads us to the conclusion that in the "true" potential, accreted GCs of one dwarf galaxy are not on similar orbits but have a distribution function that is more complex. 

\begin{figure}[htbp]
\begin{center}
 \includegraphics[width=3.3in]{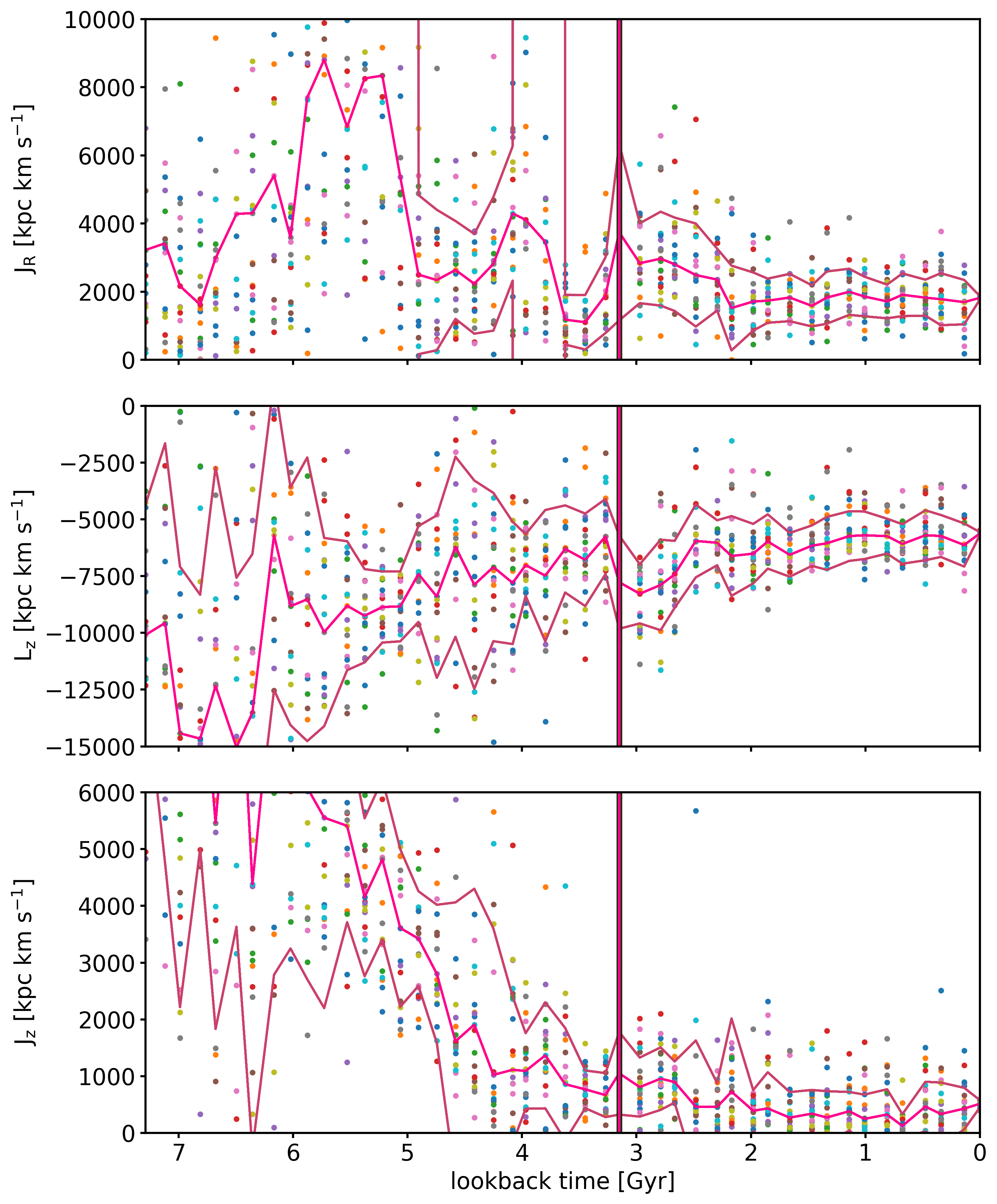} 
 \caption{Action evolution of particles of prog2 which are found in a small box in action space at $t=0$, indicating they move then on a similar orbit. The time of the merger is shown by the vertical line and mean and standard deviations by the other lines. Each colored point represents a GC. They do not stay on constant orbits and their proximity at $t=0$ seems to be only by chance.}
 \label{fig3}
\end{center}
\end{figure}

{\underline{\it Test: evolution of GCs on same orbit at $t = 0$}}. Now we consider GCs whose actions are close together in the last simulation snapshot. The selection of these GCs is made by taking the mean of each action distribution and find all GCs within a cube centered on the means. We pick 21 GCs on a similar orbit. Looking at their time evolution in Figure \ref{fig3} leads to the conclusion that they are only on similar orbits at $t=0$ by chance and will have different properties soon again suggesting complex dynamical mechanisms in the evolving galaxy that would need to be taken into account in the modeling. 

\section{Implications}
{\underline{\it Be careful with adaptive dynamics}}. The assumption that merged GCs from one progenitor are on a single orbit is too simple; their actions do not clump in action space in the correct potential. On the contrary, minimizing their spread in action space might favor wrong potential parameters. Therefore, more realistic distribution functions are needed!

{\underline{\it Actions are not always constant}}. We found that in \textit{Auriga}, integrals of motion of accreted particles vary significantly with time. This could either be due to numerical artifacts in the simulation, or could imply that the build-up of MW like galaxies is complicated and even after a Hubble time not settled. The assumption of adiabatically evolving potentials (e.g. in \cite{Yang19}) might not be applicable to cosmological simulations and should also be reconsidered regarding the MW and external galaxies.


\textit{Acknowledgements} SL and GvdV acknowledge support from the European Research Council (ERC) under the European Union’s Horizon 2020 research and innovation programme under grant agreement No 724857 (Consolidator Grant ArcheoDyn).

\end{document}